# Power Law Signature in Indonesian Population
## Empirical Studies of *Kabupaten* and *Kotamadya* Population in Indonesia

Ivan Mulianta[†], Hokky Situngkir[††], Yohanes Surya[†††]

[†] Dept. Dynamical System Modeling, Bandung Fe Institute. mail: ivanm@students.bandungfe.net
[††] Dept. Computational Sociology, Bandung Fe Institute. mail: hokky@elka.ee.itb.ac.id
[†††] Senior Fellow, Surya Research Intl. mail: yohaness@centrin.net.id

**Abstract**

The paper analyzes the spreading of population in Indonesia. The spreading of population in Indonesia is clustered in two regional terms, i.e.: *kabupaten* and *kotamadya*. It is interestingly found that the rank in all *kabupaten* respect to the population does not have fat tail properties, while in the other hand; there exists power-law signature in *kotamadya*. We analyzed that this fact could be caused by the equal or similar infrastructural development in all regions; nevertheless, we also note that the first 20 *kabupatens* are dominated in Java and Sumatera. Furthermore, the fat tail character in the rank of *kotamadya* could be caused by the big gap between big cities one another, e.g.: Jakarta, Surabaya, and others. The paper ends with some suggestions of more attention to infrastructural development in eastern regional cities.

## 1. Background on Population Studies

We analyze cities in Indonesia but only concern most on respective population. In some countries, the population occasionally discussed on the cities (Gabaix, 1999; Anderson & Ge, 2003; Blank & Solomon, 2000, Henderson & Wang, 2003; Overman and Ioannides, 2000), however every country have their own criterion-defining on cities and any other regional areas to discuss the population with. For example, in the United States, an area called city when the population reached at minimum 2.5 million (U.S. Census Bureau, 2000).

Based on Indonesian law (Undang-Undang Republik Indonesia 22/1999) about regional autonomy, Indonesia region is composed by:
a) *Daerah Tingkat I* (Provinces)
b) *Daerah Tingkat II* or *Kabupaten* (Regionals)
c) *Kotamadya* (Municipals)

Daerah Tingkat I, *kabupaten*, and *kotamadya* have autonomy and without hierarchy among one another. In the rest of the paper we analyze the populations of Indonesia based on the latest two items, i.e.: *kotamadya and* kabutapen. According to the Indonesian law (Undang-Undang Republik

Indonesia 22/1999 chapter 1 section 1q), *kotamadya* is as an area, which its main activities are not agriculture and *kabupaten* is an area, which its main activities are agriculture.

According to the state governmental act (Peraturan Pemerintah 129/2000), the considerations of area to becoming kabupaten or kotamadya are:
  i) Economic ability, such as GDP
  ii) Potential sources, such as man-power, health facility, transportation facility, etc.
  iii) Socio-cultures, e.g. number of theatre buildings
  iv) Socio-politics, e.g. participation rate in general election
  v) Populations number
  vi) Area size

## 2. Basic Model
### 2.1 Mechanism of Creating Power Law
Power law is a function of the form:

$$f(x) = f(1) x^{-\tau} \qquad \ldots (1)$$

where $\tau$ is a constant. As analyzed in Situngkir & Surya (2003), power laws are known to occur at critical point on phase transition in physics. There are some circumstances emerging the power laws, e.g.: the self-organized criticality, the phase transition phenomena, and the highly optimized tolerance. In short, the discussion on the existence of power-laws in many occasions is opening the further advancement on the complex adaptive system (Per Bak, 1987).

We rank cities based on its respective population. Suppose a certain quantity *x* has a power law distribution, as in equation (1). The integral under the distribution from *x* to ∞ is:

$$R(x) = f(1) \int_{x}^{\infty} y^{-\tau} dy = \frac{f(1)}{1-\tau} x^{-\tau+1} \qquad \ldots (2)$$

This quantity is recognized as the rank. If *f(x)* is the appropriate histogram of *x*, then *R(x)* is a number of measurements that had a value greater than or equal to *x*. In other words, if we number *n* measurements from *1* (greatest) to *n* (smallest), then the number given to a measurement *x* is *R(x)*. We see that if *f(x)* is power law, then so is *R(x)*. Often people plot a so-called rank plot, i.e.: *R(x)* is plotted versus *x*. This is better than making a histogram, because it does not require us to bin the data - each data point is counted separately. In some circumstances, the rank plots can be also sometimes recognized as the cumulative distribution function (CDF).

Regarding to the Indonesian population, we recognize that the social system has the property of self-organized. People want to satisfy themselves (rather than maximize it; *see* Paul Ormerod, 2000:125). This fact leads to the urbanization on big cities, e.g. Jakarta, which is believed to offer more satisfactions and opportunities. Suppose there are $i_1, i_2, \ldots, i_N$ individuals live in *kabupaten* or *kotamadya*. In Indonesia, there are *M kabupaten* and *O kotamadya*. Each individual *i*, has experiental spaces of $s_i$, thus the total experiental space is $S = \{s_1, s_2, \ldots, s_N\}$, representing knowledge about places to live in. Decision of each individual $i$ in time steps $t$, $D_i^t = \{d_1^t,...,d_N^t\}$, could be whether to stay or to migrate. The decision is much influenced by the pay-off function as the function within experiental spaces and the decisions, $\omega_i^t = f_i(S_i, d_i^t)$. Individual decides whether to live or to migrate in order to be satisfied, or mathematically $d_i^t = \max \omega_i^t$. If individual $i$ is not satisfied when living in *kabupaten* or *kotamadya*, he organize himself to migrate to another *kabupaten* or *kotamadya* until find the most satisfaction.

Let $P_{kabupaten} = \{I_1, I_2, ..., I_M\}$ is a set of populations living in $m$ *kabupaten* and they are ranked based on number of populations, index *1* for the biggest, *M* for the smallest. For *kotamadya*, we denote $P_{kotamadya} = \{I_1, I_2, ..., I_O\}$, index *1* for the biggest, *O* for the smallest.

**2.2 Analysis of Indonesian Population Data**

Now, we analyze Indonesian cities population to see how many populations live in Indonesian cities. Here, we make a rank of *kabupaten* and *kotamadya* according to its population; the biggest population gets rank 1, the second biggest gets rank 2, and so on until the smallest. In this case, we will consider Jakarta as *kotamadya* since its main activities are non-agriculture. We make $P_{(1)}$ the biggest, $P_{(2)}$ the second, and so on, to have:

$$P_{(r)} \approx r^{-\frac{1}{\alpha}} \quad \ldots (3)$$

$$P_{(r)} = P_1 \, r^{-\alpha}$$

It is interesting that the rank plots of *kabupaten* do not show power law signature (see figure 1). Since it does not have the fat tail properties, we can consider it to be Gaussian in every census year (with α ≈ 0.5), summarized in Table 1.

**Table 1**
α coefficient in each census year

| Census Year | α |
|---|---|
| 1961 | 0.5968 |
| 1971 | 0.4981 |
| 1980 | 0.5239 |
| 1990 | 0.598 |
| 2000 | 0.5706 |

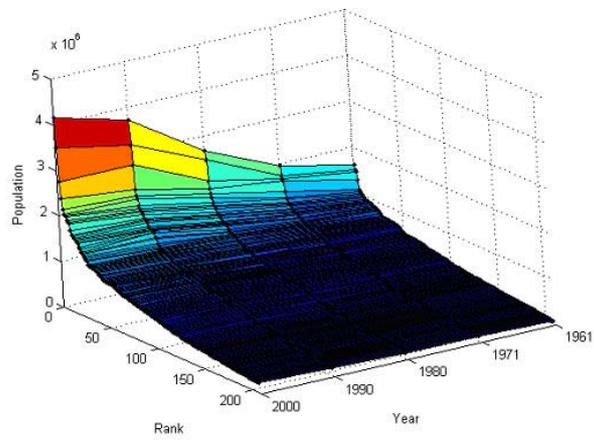

**Figure 1**
Populations living in *kabupate*n per census year

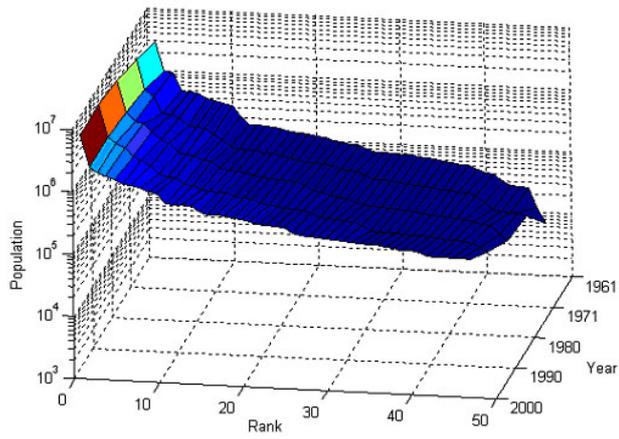

**Figure 2**
Populations living in *kotamady*a per census year

In figure 2, we see that the ranked cities show power law signature. The coefficient *α* rises in the period of 1961-1990, but in 2000, it is below *1*. For all α in each year, we summarized in Table 2.

**Table 2**
α coefficient in each census year

| Census Year | α |
|---|---|
| 1961 | 1.107 |
| 1971 | 1.14 |
| 1980 | 1.227 |
| 1990 | 1.072 |
| 2000 | 0.9882 |

From Figure 1, we can say that most populations in *kabupatens* are not only concentrated in few *kabupatens* only. It is contrasted to the result we have in *kotamadya* showing the power law signature. It means that most populations live in few *kotamadyas* only, i.e.: Jakarta, Surabaya, and Bandung and only small populations live in the rest of *kotamadyas*. Thus, we can say there is a big gap among *kotamadyas* in Indonesia. This probably comes from the gap of infrastructural development becoming the major factor that makes people migrate to Jakarta, Surabaya, and other major big cities. The gap, however, turns out to be the critical points among people that self-organizing to have better living. That is why we can recognize the self-organizing in critical points among people become the major reason of the emerging power-laws among cities in Indonesia.

Furthermore, beside economic reasons, there could be another reason explaining why people urbanize, e.g.: city attraction on the lifestyles within (see: Louis Wirth, 1938; George Simmel, 1971; Claude Fischer, 1975), the major development on transportation sector ease people (see: D.F Batten, 1995), or a myth that in cities there are many job opportunities. In the case of Indonesia, it is often heard that urbanized people were asked by their relatives to help running their business in cities.

## 3. Discussions

To see dynamics of population in more micro view, we plot *kabupaten* or *kotamadya* on its rank (increase or decrease), population, in respective years (from 1961 to 2000).

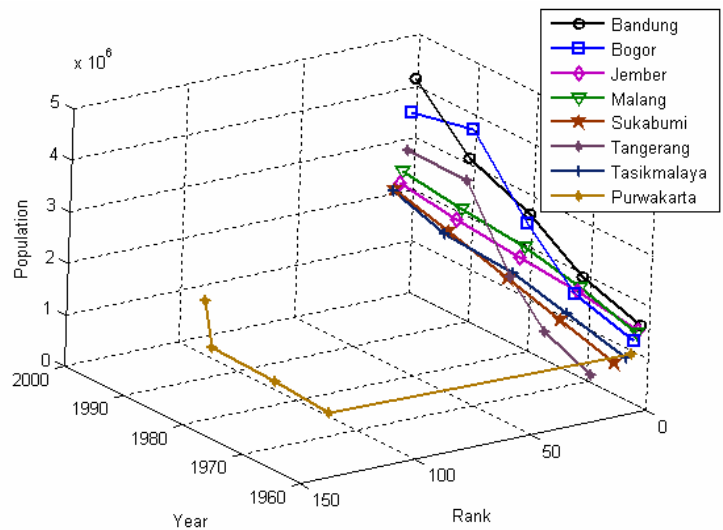

**Figure 3**
Population Dynamics of *Kabupaten*

In figure 3, *Kabupaten* Bandung seems to dominate the 1st rank. We take *Kabupaten* Tangerang and *Kabupaten* Purwakarta as examples because they show fascinating phenomenon. *Kabupaten* Purwakarta in 1961 was the 6th, but in period 1971-2000, its rank dropped to around 80's. It might be caused by the division of *Kabupaten* Purwakarta became two i.e.: *Kabupaten* Subang and *Kabupaten* Purwakarta; before the separation, most population lived in Subang. Another possible explanation is Purwakarta is a regional connecting three big cities (Jakarta, Bandung, and Cirebon). Most lands in Purwakarta are used for farming, cultivation, and forestry. Instead of becoming farmer, people prefer to urbanize to Jakarta, Bandung, and Cirebon in order to seek another job opportunities. In other case, *Kabupaten* Tangerang, its ranks always hike in the period of 1961-2000, the main factor was the opening of many industries, such as shoes, fashions, *et cetera*. The openings of those industries attract many people to migrate.

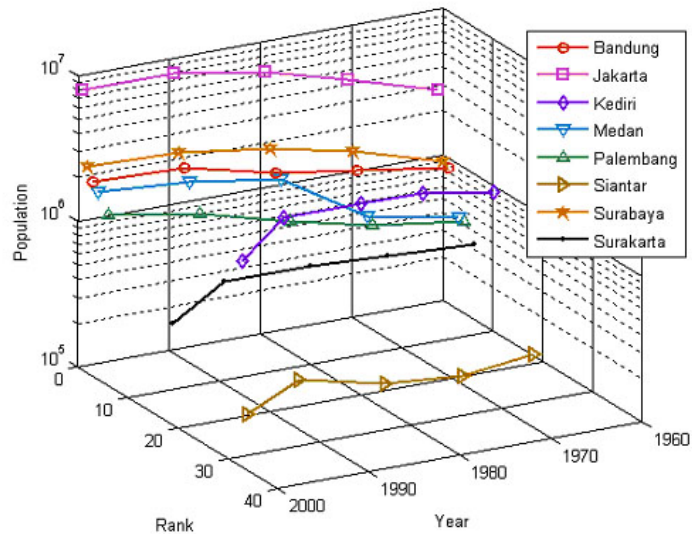

**Figure 4**
Population Dynamics of *Kotamadya*

In figure 4, we can see that big-three cities always hold the same positions from 1961 to 2000. None gives interesting pattern compared to *Kabupaten* Purwakarta and Tangerang. The big gap of development in Jakarta and other *kotamadya* leads people to urbanize to Jakarta, for it is the so called fact that people looking better opportunities that can only be offered by Jakarta.

How is the spreading of population among cities in Indonesia according to the facts of decentralized and centralized governance? We can see this property by plotting the evolution of cities normalized to Jakarta respect to the distance. Let Jakarta be a center of Indonesia ($d=0$), and we plot the cities (*kotamadya*) then plot the population according to its distance $d$.

As showed in figure 5, we may divide the distances ($d$) from Jakarta to several *kotamadya* into three emerging groups, i.e.:

    A.   0 < d < 500 miles
    B.   500 < d < 1000 miles
    C.   > 1000 miles

*Kotamadya* in-group A (*0<d<500 miles*) shows the decreasing trend (1961-2000), e.g.: Bandung, Surabaya. It is convenient to hypothesize that many people urbanized to Jakarta looking for better living and opportunities, while the transportation facilities from Bandung, Surabaya to Jakarta ease people to urbanize. If we look back to our assumption, the self-organized criticality, then it is clear that Jakarta offer more attractions and opportunities on most people to migrate, $d_i^t = \max \omega_i^t$. *Kotamadya* in group-B (*500<d<1000 miles*) is more interesting; some *kotamadya'* populations decrease, while in some places increase, e.g.: Medan. Medan is known as the biggest *kotamadya* in the island of Sumatera, and the economic activities are better than in others in the whole big island. In short, we can say that Medan is a major city in Sumatera and most people in Sumatera prefer to migrate to Medan than Jakarta because of the transportation cost is less while in the same time the culture in Medan is more related and similar to them. The *kotamadya* in group C (*d>1000 miles*); the trend of normalized population is relatively constant. We cannot justify that population in-group C are happy to stay there or the development is magnificent. It is much sound if we think that people have to spend expensive cost to migrate to Jakarta while not many people can do that – based on the minimum transportation development in these areas. We can say, in *kotamadya* C, there is local $\max \omega_i^t$ in each *kotamadya* that does not drive people to migrate to inter-*kotamadya* iii or to Jakarta. Interestingly, we can see that most cities in the group C is in the eastern region of Indonesia – a point showing the lack of infrastructural development in the period.

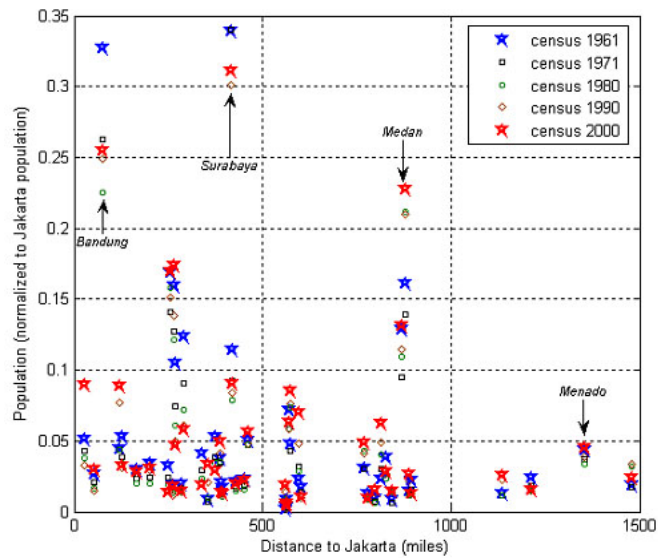

**Figure 5**
The distance and the populations living in several *kotamadya*

## 4. Concluding Remarks

We show that power law signature in *kotamadyas* of Indonesia that assumed to be caused by the self-organized criticality among people. People decide whether to stay or to migrate in order to be satisfied. If people live in *kotamadya*, but they are not satisfied, they would organize to migrate to another *kotamadya* offering better living according to their experiential spaces. In this case, the self-organization tends to place the critical situations among people to migrate or to stay to have a better living.

The population ranked in *kabupaten*-manner does not have fat tail properties, contrasted to the *kotamadya,* the power-law exists. The gaussian of the rank of *kabupatens* shows that the major problem is not among regional but among cities. The power-law signature among *kotamadya* can be understood in accordance of big gap development amongst; believed to have better living, people urbanize to Jakarta. There is no way to stop the flow of the urbanization, since the attraction of big city will always promise better living. Regarding the discussions in the paper, we suggest that the government can break the migration flow by paying attention to the infrastructural development in other (remote) *kotamadyas*. Probably, a good project on this is the project of accelerating development in Eastern Indonesia.

## Further Works

In further works, we want to make agent based modeling of regional mobility and poverty trap connected with population distribution in Indonesia. By looking at the micro level (agent) behavior, we want to see the result in macro level as aggregate. By doing this, we could make proper suggestion technical policies.

## Acknowledgement

The authors thank the Surya Research International for financial support and Badan Pusat statistik Indonesia (Indonesian Central Bureau of Statistics). Ivan Mulianta thanks Sarityastuti and Yakobus Sitompul for gathering data. All faults remain authors'.